\documentclass[amsthm]{elsart}
\usepackage{amsfonts}
\usepackage{amsmath}
\usepackage{graphicx}
\newcommand{\p}{^{\prime}}
\renewcommand{\d}{\,\mathrm{d}}

\begin{document}
\begin{frontmatter}
\title{K. Schwarzschild's problem in radiation transfer theory}
\runtitle{K. Schwarzschild's problem in transfer theory}
\author [Lyon]{B. Rutily\thanksref{coresp},}
\author [Lyon] {L. Chevallier,}
\author [JP] {J. Pelkowski}

\address [Lyon]{Centre de Recherche Astronomique de Lyon (UMR 5574 du CNRS),
Observatoire de Lyon, 9 avenue Charles Andr\'{e}, 69561 Saint-Genis-Laval Cedex, France}
\address [JP] {Institut f\"ur Meteorologie und Geophysik, J.W. Goethe Universit\"at Frankfurt, Robert Mayer Strasse 1, 
D-60325 Frankfurt a.M., Germany}

\thanks[coresp]{Corresponding author. Tel.:+33-478-868-379; Fax.:+33-478-868-386.\\
\textit{Email address:} \texttt{rutily@obs.univ-lyon1.fr} (B. Rutily).}
\runauthor{Rutily, Chevallier and Pelkowski}

\date{8 September 2004}
\received{8 September 2004}
\accepted{16 May 2005}

\begin{abstract}
We solve exactly the problem of a finite slab receiving an isotropic radiation on one side and no radiation 
on the other side. This problem - to be more precise the calculation of the source function within the slab - 
was first formulated by K. Schwarzschild in 1914. We first solve it for unspecified albedos and optical 
thicknesses of the atmosphere, in particular for an albedo very close to 1 and a very large optical thickness 
in view of some astrophysical applications. Then we focus on the conservative case (albedo = 1), which is of 
great interest for the modeling of grey atmospheres in radiative equilibrium. Ten-figure tables of the 
conservative source function are given. From the analytical expression of this function, we deduce 1) a simple 
relation between the effective temperature of a grey atmosphere in radiative equilibrium and the temperature of 
the black body that irradiates it, 2) the temperature at any point of the atmosphere when it is in local 
thermodynamical equilibrium. This temperature distribution is the counterpart, for a finite slab, of Hopf's 
distribution in a half-space. Its graphical representation is given for various optical thicknesses of the atmosphere. 
\end{abstract}
\end{frontmatter}

\maketitle

\section{Introduction}\label{sec1}
In a paper considered as foundational in transfer theory, K. Schwarzschild \cite{schwarzschild1914} modeled the 
solar atmosphere as a finite slab irradiated on one side by black body radiation. He set up, for the first time, the 
complete radiation transfer equation - including the scattering integral - and inferred the following integral 
equation for the source function $S$:
\begin{equation}
\label{eq1}
S(\tau) = S_0 (\tau) + \frac{a}{2} \int_{0}^{b} E_{1} (| \tau - \tau\p |) S(\tau\p) \d\tau\p,
\end{equation}
with
\begin{equation}
\label{eq2}
S_0(\tau) = (1-a)B^*(\tau) + \frac{a}{2} B(T_b)E_2(b-\tau).
\end{equation}
The first term on the right-hand side of Eq. (\ref{eq1}) describes the contribution of the (internal and external) sources, 
and the integral term is the scattering term. $a \in [0,1]$ is the albedo (supposed constant) for single scattering by a 
volume element, $b>0$ the optical thickness of the atmosphere, and $\tau \in [0, b]$ is the optical depth variable. Eq. (2) 
shows that the source term $S_0$ includes the thermal emission $B^*$ of the atmosphere, as well as the contribution of the 
external sources. In our model, they will be prescribed as black body radiation of temperature $T_b$, incident on the surface
 $\tau = b$. $B(T_b)$ is the Planck function at the temperature $T_b$. There is no radiation incident on the other boundary
 surface $\tau=0$. The scattering of light is supposed monochromatic and isotropic, which assumption gives rise to the 
exponential integral functions
\begin{equation}
\label{eq3}
E_{n} (\tau) = \int_{0}^{1} \exp (-\tau / u) u^{n-2}\d u \qquad (n \geq 1)
\end{equation}
in the kernel and the free term of Eq. (\ref{eq1}). Frequency dependence of quantities will not be made explicit.\\
The integral form of the radiation transfer equation was established before the publication of
Schwarzschild's paper, by Lommel \cite{lommel1889}, Chwolson \cite{chwolson1890} and King \cite{king1913}. In these
articles, Eq. (\ref{eq1}) with $S_0(\tau) = \exp(-\tau/\mu)$ ($\mu>0$) was derived directly, with no reference to the 
transfer equation. Integral equations of the form (\ref{eq1}) have been studied intensively since these pioneering works. 
Three general methods for solving them, which apply to the particular source term (\ref{eq2}), are described in 
\cite{rutily-bergeat1994}. Since the problem (\ref{eq1})-(\ref{eq2}) is linear, its solution can be written as
\begin{equation}
\label{eq4}
S(\tau) = (1-a)\int_{0}^{b} R(a,b,\tau, \tau\p)B^*(\tau\p) \d \tau\p + B(T_b)\xi_{0}(a, b,b-\tau),
\end{equation}
where $R$ denotes the resolvent kernel of the integral equation (\ref{eq1}) and $\xi_0$ the solution to the following 
integral equation:
\begin{equation}
\label{eq5}
\xi_0(a,b,\tau) = \frac{a}{2}E_2(\tau) + \frac{a}{2} \int_{0}^{b} E_{1} (| \tau - \tau\p |) \xi_0(a,b,\tau\p) \d\tau\p.
\end{equation}
In view of evaluating the integral on the right-hand side of Eq. (\ref{eq4}), the model under investigation must be further
specified to make explicit the $\tau$-dependence of the function $B^*$. When the distribution of internal sources 
is uniform in the slab, $B^*$ is independent of $\tau$ and then the thermal emission of the atmosphere is characterized by 
the function
\begin{equation}
\label{eq6}
Q(a,b,\tau) = \int_{0}^{b} R(a,b,\tau, \tau\p) \d \tau\p .
\end{equation}
In the conservative case ($a=1$), both functions $\xi_0$ and $Q$ were studied in detail by Heaslet and Warming 
\cite{heaslet-warming1965}-\cite{heaslet-warming1967} in view of some applications in heat transport theory. Their functions 
$\Theta(\tau)$ and $\Theta_s(\tau)$ coincide with our functions $\xi_0(1,b,\tau)$ and $(1/4)Q(1,b,\tau)$, respectively. In the 
non-conservative case ($0\leq a<1$), the function $\xi_0$ is briefly evoked in Section 8 of \cite{heaslet-warming1968}, 
while the literature on the function $Q$ is mentioned in a recent article devoted to this function \cite{chevallier-rutily2003}.\\
In what follows, we will be interested exclusively in the second term on the right-hand side of Eq. (\ref{eq4}), which describes the 
contribution of the external sources. The exact expression for the $\xi_0$-function is given for an unspecified albedo $a$, 
including $a=1$. This particular case is of great importance for our understanding of grey atmospheres in radiative 
equilibrium. Actually, it is known \cite{hopf1934} that the solution to Schwarzschild's problem with $a=1$ is the 
source function, integrated over frequency, of a grey atmosphere in radiative equilibrium. It thus yields the temperature 
of the atmosphere if the latter may be assumed to be in local thermodynamical equilibrium. This particular problem has 
been solved exactly in a half-space, that is, for $a=1$, $b=+\infty$ and $S_0=0$ (since the sources are at infinity). The 
solution is due to Hopf \cite{hopf1934}. In a finite slab ($b<+\infty$), the conservative problem (\ref{eq1})-(\ref{eq2}) is equivalent to 
(\ref{eq5}) with $a=1$. Yamamoto \cite{yamamoto1956} failed to solve it exactly, as remarked by Sobolev \cite{sobolev1962}. 
Heaslet and Warming's solution \cite{heaslet-warming1965}-\cite{heaslet-warming1967} is approximate, in contrast with that 
given by Kriese and Siewert \cite{kriese-siewert1970}, who used the Case method. Four-figure tables of the conservative functions $\xi_0$ and $Q$ are given in \cite{heaslet-warming1967}, in good agreement 
with the six-figure tables of \cite{kriese-siewert1970}.\\
The outline of this article is as follows: in Section 2, a recent work dealing with a problem connected with Schwarzschild's 
problem is summarized. Sections 3 to 5 are devoted to the calculation of the $\xi_0$-function in the general case, in a 
conservative atmosphere and in a semi-infinite atmosphere, respectively. The numerical evaluation of the $\xi_0$-function is
tackled in Section 6, which contains a table of this function for $a=1$ and different values of $b$. In Section 7, an 
application of the conservative case to grey atmospheres in radiative equilibrium is investigated. From the exact 
expression of the source function integrated over frequency, we deduce a simple relation between the effective 
temperature of the atmosphere and the temperature of the black body that irradiates it. We also give the temperature
of the atmosphere when it is in local thermodynamical equilibrium, a calculation which generalizes Hopf's calculation in 
a half-space to a finite slab.

\section{The standard transfer problem in a stellar atmosphere: a summary}\label{sec2}
In the simplest model atmosphere that can be conceived - a homogeneous and isothermal slab in local thermodynamical
equilibrium with no incident radiation - the key function is the function $Q$, which is the solution to the following 
integral equation
\begin{equation}
\label{eq7}
Q(a,b,\tau) = 1 + \frac{a}{2} \int_{0}^{b} E_{1} (| \tau - \tau\p |) Q(a,b, \tau\p) \d\tau\p.
\end{equation}
It is clear that this function is symmetric about $\tau=b/2$, i.e., $Q(a,b,\tau)=Q(a,b,b-\tau)$. A detailed, recent study 
of this classical auxiliary function can be found in Ref. \cite{chevallier-rutily2003}, which we summarize hereafter. 
The starting point is Sobolev's expression 
\cite{sobolev1958}
\begin{equation}
\label{eq8}
Q (a,b, \tau) = \psi (a, b, b) [ \psi (a, b, \tau) + \psi (a, b, b - \tau) - \psi (a, b, b) ],
\end{equation}
which involves the function
\begin{equation}
\label{eq9}
\psi (a, b, \tau) = 1 + \int_{0}^{\tau} \phi (a, b, \tau\p) \d \tau\p.
\end{equation}
Here, $\phi= \phi(a,b, \tau)$ is the resolvent function of the problem (\ref{eq1}), which solves it for the particular free term 
$S_0(\tau)= (a/2)E_1(\tau)$. From the well-known analytical expression of this function, we derived in \cite{chevallier-rutily2003}  
the following expression of the $\psi$-function:
\begin{multline}
\label{eq10}
\psi(a,b,\tau)= 
\frac{\tilde{\alpha}_0(a,b)}{1-a}\\
 -\!R(a)\{[1\!-\xi_X(a,b,1/k(a))]\exp[-k(a)\tau]+\xi_Y(a,b,1/k(a))\exp[-k(a)(b\!-\!\tau)]\}\\
 -\frac{a}{2}\int_{0}^{1}g(a,u)\left\{[1-\xi_X(a,b,u)]\exp(-\tau/u) 
 + \xi_Y(a,b,u)\exp[-(b-\tau)/u]\right\}\d u.
\end{multline}
Let us briefly point out the definition of the coefficients and functions appearing in this expression, referring to 
\cite{chevallier-rutily2003} for details:\\
1) For $0<a<1$, the coefficient $k(a)$ is the unique root, in $]0,1[$, of the characteristic equation
\begin{equation}
\label{eq11}
1-\frac{a}{2}\frac {1}{k(a)}\ln \left[ \frac {1+k(a)}{1-k(a)} \right] = 0 ,
\end{equation}
so that $k(0)=1$ and $k(1)=0$ by continuity.\\
2) We have
\begin{equation}
\label{eq12}
R(a) = \frac{1 - k^{2}(a)}{k^{2}(a)+a-1} \quad(0<a<1). 
\end{equation}
3) For any $x\not=-1$
\begin{equation}
\label{eq13}
\xi_X (a, b, x) = \frac{a}{2}\, x \int_{0}^{1} X(a, b, u) \frac{\d u}{u + x}, 
\end{equation}
\begin{equation}
\label{eq14}
\xi_Y (a, b, x) = \frac{a}{2}\, x \int_{0}^{1} Y(a, b, u) \frac{\d u}{u + x},
\end{equation}
where $X$ and $Y$ are the classical functions of finite, plane-parallel media. Note that the integrals are 
Cauchy principal values when $x \in ]-1, 0[$.\\
4) We introduce the moments of order $n\geq 0$ of the $X$- and $Y$-functions
\begin{equation}
\label{eq15}
\alpha_n(a,b) = \int_{0}^{1}X(a,b,u)u^n \d u\quad,\quad \beta_n(a,b)=\int _{0}^{1}Y(a,b,u) u^n \d u,
\end{equation}
and we set
\begin{equation}
\label{eq16}
\tilde{\alpha}_0(a,b)= 1 - \frac{a}{2}\alpha_0(a,b)\quad,\quad \tilde{\beta}_0(a,b)= \frac{a}{2}\beta_0(a,b).
\end{equation}
These coefficients are not independent, since they satisfy
\begin{equation}
\label{eq17}
[\tilde{\alpha}_0(a,b)-\tilde{\beta}_0(a,b)][\tilde{\alpha}_0(a,b)+\tilde{\beta}_0(a,b)] = 1-a.
\end{equation}
5) Finally, the integral in the right-hand side of Eq. (\ref{eq10}) contains the function 
\begin{equation}
\label{eq18}
g(a, u) = \frac{1}{T^{2} (a, u) + [(\pi/2) a u]^{2}} \qquad (0 \leq u < 1),
\end{equation}
where
\begin{equation}
\label{eq19}
T(a,u) = 1 - \frac{a}{2} u \ln \left(\frac{1+u}{1- u}\right) .
\end{equation}
The boundary values of the $\psi$-function are \cite{chevallier-rutily2003}  
\begin{align}
\psi (a, b, 0) &= 1,\label{eq20}\\
\psi (a, b, b) &= X (a, b, \infty),\label{eq21}
\end{align}
where
\begin{equation}
\label{eq22}
X(a, b,\infty) = \frac{1}{1-a}\left[ \tilde{\alpha}_{0}(a,b) - \tilde{\beta}_{0}(a,  b) \right]=
\frac{1}{\tilde{\alpha_{0}}(a,b) + \tilde{\beta}_{0}(a,b)} 
\end{equation}
is the value at infinity of the $X$-function; whence the boundary values of the $Q$-function: 
\begin{equation}
\label{eq23}
Q (a, b, 0) = Q (a, b, b) = X (a, b, \infty).
\end{equation}
The expression (\ref{eq10}) for $\psi(a,b,\tau)$ contains the indeterminate form $\infty-\infty$ when $a \rightarrow 1$, since
$\tilde{\alpha}_0(a,b)$ is bounded and $R(a)$ diverges like $1/(1-a)$. In \cite{chevallier-rutily2003}, the functions $F_{\pm}$ 
were introduced to circumvent this difficulty. Another approach would have been to consider that the function $\psi$ is the basic 
function of the problem and to remove the indetermination analytically. This can be done by remarking that $\psi(a, b, 0)=1$,
and thus we have
\begin{align}
\label{eq24}
1 = &\frac{\tilde{\alpha}_0(a,b)}{1-a} - R(a)\{[1-\xi_X(a,b,1/k(a))]+\xi_Y(a,b,1/k(a))\exp[-k(a)b]\} \nonumber \\
 &-\frac{a}{2}\int_{0}^{1}g(a,u)\left\{[1-\xi_X(a,b,u)]+\xi_Y(a,b,u)\exp(-b/u)\right\}\d u.
\end{align}
Subtracting Eq. (\ref{eq24}) from Eq. (\ref{eq10}), one obtains the following alternative expression for $\psi$:
\begin{multline}
\label{eq25}
\psi(a,b,\tau)= 1+k(a)R(a) \tau \left\lbrace [1-\xi_X(a,b,1/k(a))]\gamma^*[1,k(a)\tau] \phantom{\sum}\right.\\
\hspace{10em}\left. \phantom{\sum} -\xi_Y(a,b,1/k(a))\gamma^*[1,-k(a)\tau]\exp[-k(a)b] \right\rbrace\\
\hspace{-2em}+ \tau\frac{a}{2}\int_{0}^{1}g(a,u)\left\lbrace[1-\xi_X(a,b,u)]\gamma^*(1,\tau/u) \phantom{\sum}\right.\\
\left. \phantom{\sum} - \xi_Y(a,b,u)\gamma^*(1,-\tau/u)\exp(-b/u)\right\rbrace \frac{\d u}{u},
\end{multline}
where the function
\begin{equation}
\label{eq26}
\gamma^*(1,x) = \frac{1}{x} \left[ 1 - \exp({-x}) \right]
\end{equation}
belongs to a class of special functions derived from the incomplete gamma function: see Ref. ~\cite{abramowitz1970}, Section 6.5. 
We note that for $x \geq 0$, $\gamma^*(1,x)$ decreases from $\gamma^*(1,0)=1$ to $\gamma^*(1,+\infty)=0$, and that 
$\gamma^*(1,x)\sim 1/x \to 0$ as $x \to +\infty$.\\
As $a \rightarrow 1$, $k(a) \rightarrow 0$ and $\gamma^*(1,\pm k\tau)$, $\exp[-k(a)b]$ tend to 1. The first term in braces 
on the right-hand side of Eq. (\ref{eq25}) tends to $1-\xi_X(a,b,1/k(a))-\xi_Y(a,b,1/k(a))$, which is proportional to 
$k(a)$: see Section 4. We thus make appear $k^2(a)R(a)$, which remains bounded as $a \rightarrow 1$.\\ 
It is clear that the function
\begin{equation}
\label{eq27}
\tilde{\psi}(a,b,\tau)=\psi(a,b,b-\tau)-\psi(a,b,b)
\end{equation} 
can be written in a form similar to (\ref{eq25}) by subtracting from both members of Eq. (\ref{eq10}) their value for $\tau=b$. 
One obtains
\begin{multline}
\label{eq28}
\tilde{\psi}(a,b,\tau)= -k(a)R(a) \tau \left\lbrace [1-\xi_X(a,b,1/k(a))]\gamma^*[1,-k(a)\tau]\exp[-k(a)b]\phantom{\sum}\right.\\
\hspace{10em}\left. \phantom{\sum} -\xi_Y(a,b,1/k(a))\gamma^*[1,k(a)\tau] \right\rbrace\\
\hspace{-2em}- \tau\frac{a}{2}\int_{0}^{1}g(a,u)\left\lbrace[1-\xi_X(a,b,u)]\gamma^*(1,-\tau/u)\exp(-b/u) \phantom{\sum}\right.\\
\left. \phantom{\sum} - \xi_Y(a,b,u)\gamma^*(1,\tau/u)\right\rbrace \frac{\d u}{u}.
\end{multline}
It follows from Eqs. (\ref{eq27}) and (\ref{eq20})-(\ref{eq21}) that the surface values of the function $\tilde{\psi}$ are
\begin{align}
\tilde{\psi}(a,b,0) &= 0,\label{eq29}\\
\tilde{\psi}(a,b,b) &= 1-X(a,b,\infty).\label{eq30}
\end{align}
In terms of the functions $\psi$ and $\tilde\psi$, the expression (\ref{eq8}) of the $Q$-function becomes 
\begin{equation}
\label{eq31}
Q(a,b,\tau)=\psi(a,b,b)[\psi(a,b, \tau)+\tilde\psi(a,b,\tau)].
\end{equation}

\section{The calculation of $\xi_0(a,b,\tau)$}\label{sec3}
A natural approach consists in replacing $E_2(\tau)$ by its definition $E_2(\tau)=\int_{0}^{1}\exp(-\tau/u)du$ on the right-hand 
side of Eq. (\ref{eq5}). Taking advantage of the linearity of this equation, we obtain the solution in the form
\begin{equation}
\label{eq32}
\xi_0(a,b,\tau)=\frac{a}{2}\int_{0}^{1}B(a,b,\tau,u)\d u,
\end{equation}
where the function $B$ is the solution to the following integral equation: 
\begin{equation}
\label{eq33}
B(a,b,\tau,u)= \exp(-\tau/u) + \frac{a}{2} \int_{0}^{b} E_{1} (|\tau - \tau\p |) B(a,b,\tau\p,u) \d \tau\p.
\end{equation}
This function is a classical auxiliary function in plane-parallel geometry, which coincides with the $X$- and 
$Y$-functions on the two boundary surfaces \cite{busbridge1960}  
\begin{equation}
\label{eq34}
B(a,b,0,u)=X(a,b,u)\quad, \quad B(a,b,b,u)=Y(a,b,u). 
\end{equation}
Inserting these expressions into Eq.(\ref{eq32}), we derive the surface values of the $\xi_0$-function
\begin{equation}
\label{eq35}
\xi_0(a,b,0)=\frac{a}{2}\alpha_0(a,b) \quad,\quad \xi_0(a,b,b)=\frac{a}{2}\beta_0(a,b).
\end{equation}
Now derive both members of Eq. (\ref{eq5}) with respect to $\tau$, remembering that $E_2'(\tau) = -E_1(\tau)$, and 
note that the solution to Eq. (\ref{eq1}) with $S_0(\tau)=(a/2)E_1(\tau)$ is the resolvent function $\phi$. Replacing
 $\xi_0(a,b,0)$ and $\xi_0(a,b,b)$ by their expressions (\ref{eq35}), one gets
\begin{multline}
\label{eq36}
\xi_{0}\p(a,b,\tau)=\frac{a}{2} \int_{0}^{b}E_{1}(|\tau-\tau\p|)\xi_{0}\p(a,b,\tau\p)\d \tau\p\\
-\tilde\alpha_0(a,b)\frac{a}{2}E_1(\tau)- \tilde \beta_0(a,b)\frac{a}{2}E_1(b-\tau).
\end{multline}
Hence
\begin{equation}
\label{eq37}
\xi_{0}\p(a,b,\tau)=-\tilde\alpha_0(a,b)\phi(a,b,\tau)- \tilde \beta_0(a,b)\phi(a,b,b-\tau),
\end{equation}
and
\begin{equation}
\label{eq38}
\xi_{0}(a,b,\tau)=-\tilde\alpha_0(a,b)\int_{0}^{\tau}\phi(a,b, \tau\p)\d \tau\p - \tilde \beta_0(a,b)\int_{0}^{\tau}
\phi(a,b,b-\tau\p) \d \tau\p+ A.
\end{equation}
It follows from Eq. (\ref{eq35}) that the constant of integration $A$ is $(a/2)\alpha_0(a,b)$. Using the definitions
(\ref{eq9}) and (\ref{eq27}) of the functions $\psi$ and $\tilde{\psi}$, one finally obtains
\begin{equation}
\label{eq39}
\xi_{0}(a,b,\tau)=1-\tilde\alpha_0(a,b)\psi(a,b,\tau)+ \tilde \beta_0(a,b)\tilde{\psi}(a,b,\tau).
\end{equation}
As a check, the boundary expressions (\ref{eq35}) of the $\xi_{0}$-function are retrieved by means of Eqs. (\ref{eq20})-(\ref{eq21}) 
and (\ref{eq29})-(\ref{eq30}).\\
 Another approach for the calculation of the $\xi_0$-function consists in substituting in Eq. (\ref{eq32}) the expression 
for $B(a,b,\tau,u)$ given in \cite{rutily2003a}, Eqs. (66)-(68). The integration with respect to $u$ can be performed 
analytically, and the solution (\ref{eq39}) is derived by transforming the integral along the imaginary axis with the 
help of the residue theorem.\\
Introducing the escape probability function \cite{sobolev1967}
\begin{equation}
\label{eq40}
P(a,b,\tau)=1-(1-a)Q(a,b,\tau),
\end{equation}
it follows from Eqs. (\ref{eq39}), (\ref{eq27}), (\ref{eq21}), (\ref{eq22}) and (\ref{eq8}) that
\begin{equation}
\label{eq41}
\xi_{0}(a,b,\tau)+\xi_{0}(a,b,b-\tau)=P(a,b,\tau),
\end{equation}
which clarifies the link between the functions $Q$ and $\xi_{0}$. In a conservative atmosphere, $P(1,b,\tau)=1$ and the 
above equation reduces to the already known relation \cite{heaslet-warming1965}
\begin{equation}
\label{eq42}
\xi_{0}(1,b,\tau)+\xi_{0}(1,b,b-\tau)=1,
\end{equation}
which shows that the conservative $\xi_0$-function is antisymmetric about the point $\tau=b/2$, where it is equal to 1/2.\\
The probabilistic interpretation of the relation (\ref{eq42}) is clear. We remind the reader that $(1/4\pi)B(a,b,\tau,u)d\Omega$
is the probability that a photon at level $\tau$ leaves the slab, after multiple scattering, within an element of solid angle 
$d \Omega$ around a direction defined by the angle $\arccos u$ with respect to the outward normal at the top of the atmosphere. From Eq. 
(\ref{eq32}), we deduce that $\xi_0(a,b,\tau)$ is the probability of emergence, through the upper boundary plane, of a photon 
incident on a particle placed at level $\tau$. $\xi_0(a,b,b-\tau)$ is the probability of emergence from the lower boundary plane, 
and the sum of these two functions is the probability of emergence $P(a,b,\tau)$.  

\section{The case $a \rightarrow 1$}\label{sec4}
The numerical calculation of $\xi_{0}(a,b,\tau)$ with the help of Eqs. (\ref{eq39}), (\ref{eq25}) and (\ref{eq28}) is easy as 
long as the albedo $a$ is not too close to 1 and the optical thickness $b$ is large enough to make sure that the product $k(a)b$ 
is greater than 30. When $k(a)b\leq30$, a difficulty arises when computing the functions $\psi$ and $\tilde \psi$, since the product $k(a)R(a)$, which
appears in the right-hand side of Eqs. (\ref{eq25}) and (\ref{eq28}), diverges like $1/k(a)$ as $a \rightarrow 1$. Since it 
is multiplied by a quantity which tends to be proportional to $k(a)$ as $a \rightarrow 1$, an indeterminate form $\infty \times 0$
appears when $a \rightarrow 1$, $k(a) \rightarrow 0$. To remove it analytically, we write
\begin{equation}
\label{eq43}
\tilde{\alpha}_{0}\psi - \tilde{\beta}_{0} \tilde{\psi}= \frac{1}{2}(\tilde{\alpha}_{0}+\tilde{\beta}_{0})(\psi-\tilde{\psi})
+ \frac{1}{2}(\tilde{\alpha}_{0}-\tilde{\beta}_{0})(\psi+\tilde{\psi})
\end{equation}
with which Eq. (\ref{eq39}) becomes
\begin{equation}
\label{eq44}
\xi_{0}(a,b,\tau)=1-\frac {1}{2}[\tilde\alpha_{0,+}(a,b)\psi_{-}(a,b,\tau)+ \tilde \alpha_{0,-}(a,b)\psi_{+}(a,b,\tau)],
\end{equation}
where
\begin{equation}
\label{eq45}
\tilde{\alpha}_{0,\pm}(a,b) = \tilde{\alpha}_{0}(a,b) \pm \tilde{\beta}_{0}(a,b),   
\end{equation}
and
\begin{equation}
\label{eq46}
\psi_{\pm}(a,b, \tau) = \psi(a,b,\tau) \pm \tilde{\psi}(a,b,\tau).   
\end{equation}
From Eqs. (\ref{eq25}), (\ref{eq28}) and the identity
\begin{equation}
\label{eq47}
\gamma^{*}[1,k(a)\tau]\pm\gamma^{*}[1,-k(a)\tau]\exp[-k(a)b]= \gamma^{*}[1,k(a)\tau]\lbrace 1\pm \exp[-k(a)(b-\tau)]\rbrace,
\end{equation}
the expression for $\psi_{\pm}(a,b,\tau)$ is seen to be
\begin{multline}
\label{eq48}
\psi_{\pm}(a,b,\tau)= 1+k(a)R(a) \tau [1-\xi_X(a,b,1/k(a))\pm\xi_Y(a,b,1/k(a))]\\
\hspace{12em}\times\gamma^*[1,k(a)\tau]\lbrace1\mp\exp[-k(a)(b-\tau)]\rbrace\\
+ \tau\frac{a}{2}\int_{0}^{1}g(a,u)[1-\xi_X(a,b,u)\pm\xi_Y(a,b,u)]\gamma^*(1,\tau/u)\lbrace1\mp\exp[-(b-\tau)]\rbrace \frac{\d u}{u}.
\end{multline}
The integral term in this expression involves the functions $\xi_X$ and $\xi_Y$, which are difficult to compute on $[0,1]$. 
As in Section 5 of \cite{chevallier-rutily2003}, we utilize the classical $H$-function and the functions $\zeta_{\pm}$, 
first introduced in Ref. \cite{rutily2003b}, which makes it possible to calculate the functions $X$, $Y$, $\xi_X$ and $\xi_Y$
 in one step. The algorithm to calculate them is summarized in Section 6.3 of \cite{chevallier-rutily2003}. 
Formulas expressing the functions $\xi_X$ and $\xi_Y$ are, for $x \geq 0$,
\begin{align}
1 - \xi_X (a, b, x) = \frac{1}{H(a, x)} \frac{1}{2} (\zeta_{+} + \zeta_{-})(a, b, x), \label{eq49} \\
\xi_Y (a, b, x) = \frac{1}{H(a, x)} \frac{1}{2} (\zeta_{+} - \zeta_{-}) (a, b, x). \label{eq50}
\end{align}
Consequently,
\begin{equation}
\label{eq51}
1 - \xi_X (a, b, x) \pm \xi_Y (a, b, x) = \frac{1}{H(a, x)} \zeta_{\pm} (a, b, x),
\end{equation}
and the expression (\ref{eq48}) for $\psi_{\pm}(a,b,\tau)$ now reads
\begin{multline}
\label{eq52}
\psi_{\pm}(a,b,\tau)= 1+\tau \left\lbrace\frac{k(a)R(a)}{H(a,1/k(a))}\zeta_{\pm}(a,b,1/k(a))\right.\\
\left.\qquad\qquad\times\gamma^*[1,k(a)\tau]\lbrace 1\mp\exp[-k(a)(b-\tau)]\rbrace\right.\\
\left. + \frac{a}{2}\int_{0}^{1}g(a,u)\zeta_{\pm}(a,b,u)\gamma^*(1,\tau/u)\lbrace1\mp\exp[-(b-\tau)]\rbrace \frac{\d u}{u} \right\rbrace.
\end{multline}
Since $1-\exp[-k(b-\tau)]= k(b-\tau)\gamma^*[1, k(b-\tau)]$, we have
\begin{multline}
\label{eq53}
\psi_{+}(a,b,\tau)= 1+\tau(b-\tau)\left\lbrace\frac{k(a)R(a)}{H(a, 1/k(a))}k(a)\zeta_{+}(a,b,1/k(a))\right.\\
\left.\hspace{10em}\times\gamma^*[1,k(a)\tau]\gamma^*[1,k(a)(b-\tau)]\right.\\
\left.+ \frac{a}{2}\int_{0}^{1}g(a,u)\zeta_{+}(a,b,u)\gamma^*(1,\tau/u) \gamma^*[1,(b-\tau)/u]\frac{\d u}{u^2}\right\rbrace,
\end{multline}
and
\begin{multline}
\label{eq54}
\psi_{-}(a,b,\tau)= 1+\tau \left\lbrace\frac{k(a)R(a)}{H(a,1/k(a))}\zeta_{-}(a,b,1/k(a))\right.\\
\left.\qquad\qquad\qquad\times\gamma^*[1,k(a)\tau]\lbrace 1+\exp[-k(a)(b-\tau)]\rbrace\right.\\
\left.+ \frac{a}{2}\int_{0}^{1}g(a,u)\zeta_{-}(a,b,u)\gamma^*(1,\tau/u))\lbrace 1 +\exp[-(b-\tau)/u]\rbrace \frac{\d u}{u}\right\rbrace.
\end{multline}
As $a \to 1$, the coefficients $k(a)R(a)/H(a,1/k(a))$, $k(a)\zeta_{+}(a,b,1/k(a))$ and $\zeta_{-}(a,b,1/k(a))$ remain bounded, 
with the following limits \cite{chevallier-rutily2003} 
\begin{align}
\frac{k(a)R(a)}{H(a,1/k(a))} \vert_{a=1}&= \sqrt{3},\label{eq55}\\
k(a)\zeta_{+}(a,b,1/k(a))\vert_{a=1}&= \frac{\sqrt{3}}{2}\beta_0(1,b),\label{eq56}\\
\zeta_{-}(a,b,1/k(a))\vert_{a=1}&=\frac{\sqrt{3}}{4}(\alpha_1+\beta_1)(1,b).\label{eq57}
\end{align}
Moreover, asymptotic expansions (not given here) of these coefficients for $a \to 1$ can be written, which allows one to calculate them for
values of the albedo very close to 1. It is thus easy to compute the functions $\psi_{\pm}$ with the help of Eqs. (\ref{eq53})-(\ref{eq54}), 
even when $a \to 1$. For $a=1$, we have from Eqs. (\ref{eq55})-(\ref{eq57})
\begin{multline}
\label{eq58}
\psi_{+}(1,b,\tau)= 1+\tau(b-\tau)\left\lbrace\frac{3}{2}\beta_0(1,b)\right.\\
\left.+ \frac{1}{2}\int_{0}^{1}g(1,u)\zeta_{+}(1,b,u)\gamma^*(1,\tau/u)\gamma^*[1,(b-\tau)/u]\frac{\d u}{u^2}\right\rbrace,
\end{multline}
and
\begin{multline}
\label{eq59}
\psi_{-}(1,b,\tau)= 1+\tau \left\lbrace \frac{3}{2}(\alpha_1+\beta_1)(1,b)\right.\\
\left.+ \frac{1}{2}\int_{0}^{1}g(1,u)\zeta_{-}(1,b,u)\gamma^*(1,\tau/u))\lbrace 1 +\exp[-(b-\tau)/u]\rbrace \frac{\d u}{u}\right\rbrace.
\end{multline}
On the other hand, the coefficients $\alpha_0(a,b)$ and $\beta_0(a,b)$ remain bounded as $a \to 1$, with limits verifying 
$\alpha_0(1,b)+\beta_0(1,b)=2$ \cite{busbridge1960}. As a result, 
\begin{equation} 
\label{eq60}
\tilde{\alpha}_{0,-}(1,b)=0 \qquad,\qquad \tilde{\alpha}_{0,+}(1,b)=\beta_0(1,b)
\end{equation}
and Eq. (\ref{eq44}) reduces for $a=1$ to
\begin{equation}
\label{eq61}
\xi_0(1,b,\tau)=1-\frac{1}{2} \beta_0(1,b)\psi_{-}(1,b,\tau).
\end{equation}
An alternative expression given in Ref. \cite{heaslet-warming1965} is
\begin{equation}
\label{eq62}
\xi_0(1,b,\tau)=\frac{1}{2}[1+ \beta_0(1,b)F_{-}(1,b,\tau)],
\end{equation}
where $F_-(1,b,\tau)=\psi(1,b,b-\tau)-\psi(1,b,\tau)$ is the function already introduced in Ref. \cite{chevallier-rutily2003}: 
see Eq. (43), which for $a=1$ becomes
\begin{multline}
\label{eq63}
F_{-}(1,b,\tau)=\frac{3}{4}(\alpha_1+\beta_1)(1,b)(b-2\tau)\\
+\frac{1}{2}\int_0^1\frac{g(1,u)}{H(1,u)}\zeta_{-}(1,b,u)\{\exp(-\tau/u)-\exp[-(b-\tau)/u]\} \d u.
\end{multline}
To infer Eq. (\ref{eq62}) from Eq. (\ref{eq61}), just replace $\psi_-(1,b,\tau)$ by $\psi(1,b,\tau)-\tilde{\psi}(1,b,\tau)$
$=\psi(1,b,\tau)-\psi(1,b,b-\tau)+\psi(1,b,b)$ into Eq. (\ref{eq61}), and observe that $\psi(1,b,\tau)-\psi(1,b,b-\tau)=
-F_-(1,b,\tau)$ [as seen by comparing the Eqs. (\ref{eq36}) and (\ref{eq42}) of \cite{chevallier-rutily2003}] and 
$\psi(1,b,b)=X(1,b,\infty)=1/\beta_0(1,b)$ [from Eqs. (\ref{eq32}) and (\ref{eq80}) of \cite{chevallier-rutily2003}].\\
Comparing our solution with that given by Yamamoto \cite{yamamoto1956}, we note that the function $\zeta_{-}(1,b,u)$ is
missing in the integral term of Yamamoto's expression (20) for $B(\tau)=\xi_0(1,b,\tau)$. As a result, his coefficients $Q(\tau_1)$
and $L(\tau_1)$ are incorrect. Since $\zeta_{-}(1,b,u)$ tends to 1 as $b$ tends to infinity, it is clear that Yamamoto's 
formulas are a good approximation of the solution in a slab of large optical thickness. 

\section{The case $b \rightarrow +\infty$}\label{sec5}
The relation (\ref{eq39}) is appropriate for the computation of $\xi_0(a,b,\tau)$ when the values of $a$ and $b$ are such 
that $k(a)b\geq30$. The dominant terms are the first two terms on the right-hand side, while the third term is smallest. One 
can thus calculate $\xi_0(a,b,\tau)$ without any (or very small) roundoff error. For $b=+\infty$, $\beta_0(a,b)$ and 
$\tilde{\psi}(a,b,\tau)$ vanish, and Eqs. (\ref{eq39}) and (\ref{eq25}) reduce to
\begin{equation}
\label{eq64}
\xi_{0}(a,+\infty,\tau)=1-\tilde\alpha_0(a)\psi(a,+\infty,\tau),
\end{equation}
and
\begin{multline}
\label{eq65}
\psi(a,+\infty,\tau)= 1+\tau \left\lbrace\frac{k(a)R(a)}{H(a,1/k(a))} \gamma^*[1,k(a)\tau]\right.\\
\left.+ \frac{a}{2}\int_{0}^{1}\frac{g(a,u)}{H(a,u)}\gamma^*(1,\tau/u)\frac{\d u}{u}\right\rbrace\hspace{3em}
\end{multline}
respectively. Here, $\tilde{\alpha}_0(a)=1-(a/2)\alpha_0(a)$, $\alpha_0(a)$ being the zero-order moment of the $H$-function. 
It is well known that this moment is $(2/a)[1-\sqrt{1-a}]$, so that $\tilde{\alpha}_0(a)=\sqrt{1-a}$. In Eq. (\ref{eq65}), we have 
replaced $1-\xi_X(a,b,x)$ and $\xi_Y(a,b,x)$ by their limits as $b \to +\infty$, which are $1/H(a,x)$ and $0$, respectively.
Using the relation (31) of \cite{rutily2003a}
\begin{equation}
\label{eq66}
\frac{a}{2} \int_{0}^{1}\frac{g(a,u)}{H(a,u)}\d u = \frac{1}{\sqrt{1-a}} - 1 - \frac{R(a)}{H(a,1/k(a))},
\end{equation}
the above expression of $\xi_0$ can be written in the form
\begin{multline}
\label{eq67}
\xi_{0}(a,+\infty,\tau)=\sqrt{1-a}\,\left\lbrace\frac{R(a)}{H(a,1/k(a))}\exp[-k(a)\tau]\right.\\ \left.- \frac{a}{2}\int_{0}^{1}\frac{g(a,u}{H(a,u)}\exp(- \tau/u)\right\rbrace \d u,\hspace{3em}
\end{multline}
which yields for $a=1$
\begin{equation}
\label{eq68}
\xi_{0}(1,+\infty,\tau)=1.
\end{equation}
This result means that a photon travelling in a conservative, semi-infinite atmosphere eventually leaves the atmosphere after
multiple scattering.   

\section{Numerical calculations}\label{sec6}
The numerical evaluation of $\xi_0(a,b,\tau)$ from Eqs. (\ref{eq39}) or (\ref{eq44}) requires the prior calculation 
of the coefficients $\alpha_0(a,b)$ and $\beta_0(a,b)$, which are classical, and the calculation of the functions $\psi$,
$\tilde {\psi}$, $\psi_{\pm}$ with the help of Eqs. (\ref{eq25}), (\ref{eq28}), (\ref{eq46}), (\ref{eq53}), (\ref{eq54}).
The input data are the coefficient $k(a)$, the (classical) function $H(a,x)$, and the (non-classical)
functions $\zeta_{+}(a,b,x)$ and $\zeta_{-}(a,b,x)$ for $x \in [0,1]$ and $x = 1/k(a)$. Algorithms for their computation 
are described in Section 6 of \cite{chevallier-rutily2003} and will not be repeated here.\\
In the present article, we focus on the calculation of the $\xi_0$-function in the conservative case only, with the help of 
Eqs. (\ref{eq61}) or (\ref{eq62}). In a forthcoming paper, we intend to solve a problem more general than (\ref{eq5}), with
$E_n(\tau)$ in place of $E_2(\tau)$ ($n\geq 2$). Its solution $\xi_{n-2}(a,b,\tau)$ will be tabulated for a wide range of values 
of $a$, $b$, $\tau$ and different integers $n$, including $n=2$. Our interest in the conservative case is also due to its
importance for stellar atmospheric modeling, to which we shall turn in the next section.\\
We chose to calculate $\xi_0(1,b,\tau)$ from Eq. (\ref{eq62}) instead of Eq. (\ref{eq61}). The basic coefficients
are $\beta_0(1,b)$ and $\alpha_1(1,b)+\beta_1(1,b)$. They are tabulated in \cite{heaslet-warming1965} and \cite{heaslet-warming1967},
with the first reference including useful asymptotic expressions for $b\to 0$ and $b\to+\infty$. More recent tables can be found in 
Section 9.6 of \cite{vandehulst1980}, which contains many references to previous tables. In the present work, we computed them 
accurately by means of the following formulas, given here without proof:
\begin{align}
\beta_0(1, b)&= \frac{1}{\sqrt{3}}\;\frac{1}{[b/2+q(\infty)]\rho_{-,0}(1,b)+\rho_{-,1}(1,b)},\label{eq69}\\
(\alpha_1+\beta_1)(1,b)&=\frac{2}{\sqrt{3}}\rho_{-,0}(1,b),\label{eq70}
\end{align}
where
\begin{equation}
\label{eq71}
\rho_{-,n}(1, b) = \delta_{n,0}-\frac{1}{2}\int_{0}^{1}\frac{g(1,u)}{H^{2}(1,u)}\exp(-b/u)\rho_{-}(1,b,u)u^n \d u\quad (n\geq 0). 
\end{equation}
$\delta_{n,0}$ = is the Kronecker symbol and the function $\rho_{-}(1,b,u)$ is solution to the following integral 
equation:
\begin{equation}
\label{eq72}
\rho_{-} (1, b, u) = 1-\frac{1}{2}u\int_{0}^{1}\frac{g(1,v)}{H^{2}(1,v)}\exp(-b/v)\rho_{-}(1,b,v)\frac{\d v}{v+u}.  
\end{equation}
This is a Fredholm integral equation over $[0,1]$, with a regular kernel, which can be solved with great accuracy for any 
value of the parameters $a$ and $b$ \cite{chevallier-rutily2003}.\\
Values of the $\xi_0$-function are shown in Table 1 for $a=1$ and $b=0.01$, 0.1, 0.5, 1, 10 and 100. All data are given with ten 
figures, which we expect to be significant. We have checked the 6-digit tables of Kriese and Siewert \cite{kriese-siewert1970} 
and are able to confirm all corresponding figures given by these authors.
\begin{table}
\caption{Values of $\xi_0(1,b,\tau)$ for $b$ = 0.01, 0.1, 0.5, 1, 10 and 100. The $\tau$-variable is less than $b/2$ since
$\xi_{0}(1,b,b-\tau)=1-\xi_{0}(1,b,\tau)$.}
\label{tab1}
\centering
\begin{tabular}{ll}
\begin{tabular}[t]{lll}
\hline
\noalign{\smallskip}
$b$ & $\tau$ & $\xi_0(1, b, \tau)$ \\
\noalign{\smallskip}
\hline\noalign{\smallskip}
0.01  &	  0       &  0.5126129349 \\   
      &	  0.001   &  0.5097636865 \\
      &	  0.002   &  0.5072225665 \\
      &	  0.003   &  0.5047769328 \\
      &	  0.004   &  0.5023782905 \\
      &	  0.005   &  0.5000000000 \\
\noalign{\smallskip}
\hline\noalign{\smallskip}
0.1   &	  0      &  0.5710110310  \\   
      &	  0.01   &  0.5540741012  \\
      &	  0.02   &  0.5397404273  \\
      &	  0.03   &  0.5261870381  \\
      &	  0.04   &  0.5130121656  \\
      &	  0.05   &  0.5000000000 \\
\noalign{\smallskip}
\hline\noalign{\smallskip}
0.5   &	  0      & 0.6873318619 \\   
      &	  0.05   & 0.6411364778 \\
      &	  0.1    & 0.6034439404 \\
      &	  0.15   & 0.5680837612 \\
      &	  0.2    & 0.5338125198 \\
      &	  0.25   & 0.5000000000 \\
\noalign{\smallskip}
\hline
\end{tabular}
&
\begin{tabular}[t]{lll}
\hline
\noalign{\smallskip}
$b$ & $\tau$ & $\xi_0(1, b, \tau)$ \\
\noalign{\smallskip}
\hline\noalign{\smallskip}
1     &	  0     &  0.7581464585 \\   
      &	  0.1   &  0.6945631359 \\
      &	  0.2   &  0.6428723741 \\
      &	  0.3   &  0.5941703217 \\
      &	  0.4   &  0.5468089093 \\
      &	  0.5   &  0.5000000000 \\
\noalign{\smallskip}
\hline\noalign{\smallskip}
10    &	  0   &  0.9494478780 \\   
      &	  1   &  0.8512778419 \\
      &	  2   &  0.7628978526 \\
      &	  3   &  0.6751731487 \\
      &	  4   &  0.5875728676 \\
      &	  5   &  0.5000000000 \\
\noalign{\smallskip}
\hline\noalign{\smallskip}
100   &	  0    &  0.9943073833 \\   
      &	  10   &  0.8943960588 \\
      &	  20   &  0.7957970430 \\
      &	  30   &  0.6971980286 \\
      &	  40   &  0.5985990143 \\
      &	  50   &  0.5000000000 \\
\noalign{\smallskip}
\hline
\end{tabular}
\\
\end{tabular}
\end{table}

\section{Application to grey atmospheres in radiative equilibrium}\label{sec7}
The function $\xi_0$ solves Schwarzschild's problem (\ref{eq5}), especially for a conservative atmosphere ($a=1$). This 
particular case was already evoked in K. Schwarzschild's pioneering article \cite{schwarzschild1914} as an interesting 
limiting case. Since Hopf's work \cite{hopf1934}, we know that the solution of an integral equation of the form 
(\ref{eq1}) with $a=1$ also models the frequency-integrated source function of a grey atmosphere in radiative 
equilibrium. The conservation of energy in such an atmosphere reads $J(b,\tau)=S(b,\tau)$ where $J$ and $S$ are the mean
intensity and the source function, respectively, integrated over frequency. Integrating the formal solution of the transfer equation
over the angular variable, one obtains \cite{busbridge1960}
\begin{equation}
\label{eq73}
J(b,\tau)=J_0(\tau)+\frac{1}{2}\int_{0}^{b}E_1(\vert\tau-\tau\p\vert)S(b,\tau\p)\d \tau\p,
\end{equation}
where $J_0(\tau)$ denotes the mean intensity of the direct radiation due to the exterior sources, integrated over frequency.
In our problem, the exterior sources emit black body radiation of temperature $T_b$ across the boundary plane $\tau = b$,
 so that $J_0(\tau)=(1/2)B(T_b)E_2(b-\tau)$, where $B(T_b)=(\sigma/\pi)T_b^4$ ($\sigma\,=$ Stefan-Boltzmann constant). 
Accordingly, the energy condition $J(b,\tau)=S(b,\tau)$ becomes
\begin{equation}
\label{eq74}
S(b,\tau) = \frac{1}{2}B(T_b)E_2(b-\tau)+ \frac{1}{2}\int_{0}^{b}E_{1}(|\tau-\tau\p|)S(b,\tau\p)\d \tau\p,
\end{equation}
which is an integral equation of the form (\ref{eq1}) with $a=1$. Its solution is
\begin{equation}
\label{eq75}
S(b,\tau)=B(T_b)\xi_0(1,b,b-\tau)=B(T_b)[1-\xi_0(1,b,\tau)],
\end{equation}
the latter equality resulting from Eq. (\ref{eq42}).\\
From the above solution, one can derive a simple relation between the effective temperature of the atmosphere and the 
temperature of the black body that illuminates the face $\tau=b$. The effective temperature is defined by the condition
\begin{equation}
\label{eq76}
2\pi\int_{0}^{+\infty}\int_{0}^{+1}I(0,\mu,\nu)\mu \d \mu \d \nu=\sigma T_{eff}^4
\end{equation}
($\nu=$ frequency variable), this quantity being the value, at $\tau=0$, of the integrated radiative flux $F_r(\tau)=
2\pi\int_{0}^{+\infty}\int_{-1}^{+1}I(\tau,\mu,\nu)\mu \d \mu \d \nu$ since there is no radiation incident on the face 
$\tau=0$. It follows from the formal solution of the transfer equation that \cite{busbridge1960}
\begin{multline}
\label{eq77}
F_r(\tau)= 2\pi B(T_b)E_3(b-\tau)-2\pi\int_{0}^{\tau}E_2(\tau-\tau\p)S(b,\tau\p)\d \tau\p\\+ 2\pi\int_{\tau}^{b}E_2(\tau\p-\tau)S(b,\tau\p)\d \tau\p,
\end{multline}
and condition (\ref{eq76}) becomes
\begin{equation}
\label{eq78}
2\pi B(T_b)E_3(b)+ 2\pi \int_{0}^{b}E_2(\tau)S(b,\tau)\d \tau=\sigma T_{eff}^4. 
\end{equation}
Replacing in the integral $S(b,\tau)$ by its second expression (\ref{eq75}) and remarking that
\begin{align}
\int_{0}^{b}E_2(\tau)\d \tau&=\frac{1}{2}-E_3(b),\label{eq79}\\
\int_{0}^{b}E_2(\tau)\xi_0(1,b,\tau)\d \tau&=\frac{1}{2}[1-\beta_0(1,b)(\alpha_1+\beta_1)(1,b)],\label{eq80}
\end{align}
one obtains
\begin{equation}
\label{eq81}
\int_{0}^{b}E_2(\tau)S(b,\tau)\d \tau=B(\tau_b)[-E_3(b)+\frac{1}{2}\beta_0(1,b)(\alpha_1+\beta_1)(1,b)].
\end{equation}
Replacing now $\pi B(T_b)$ by $\sigma T_b^4$, Eq. (\ref{eq78}) finally leads to
\begin{equation}
\label{eq82}
\frac{T_b}{T_{eff}}= \frac{1}{[\beta_0(1,b)(\alpha_1+\beta_1)(1,b)]^{1/4}}=[\frac{b}{(\alpha_1^2-\beta_1^2)(1,b)}]^{1/4}.
\end{equation}
The last equality in Eq. (\ref{eq82}) follows from the relation $b\beta_0(1,b)=(\alpha_1-\beta_1)(1,b)$ \cite{busbridge1960}. 
As to the Eqs. (\ref{eq79}) and (\ref{eq80}), only the second one is not self-evident: it can be found in \cite{heaslet-warming1965},
Eq. (45a). But since it is not clear to us where this equation comes from in Ref. \cite{heaslet-warming1965}, it is proved in the 
Appendix.\\ 
Eq. (\ref{eq82}) expresses the link between the effective temperature of a grey atmosphere in radiative equilibrium and 
the temperature of the black body that irradiates it on the face $\tau=b$. It does not suppose that the atmosphere is in local
thermodynamical equilibrium (LTE). In Fig. 1, the ratio $T_b/T_{eff}$ is plotted as a function of $b$ for $0\leq b\leq 100$. 
This ratio is equal to 1 for $b=0$ [since $\beta_0(1,0)=(\alpha_1+\beta_1)(1,0)=1$], and it is close to $(3b/4)^{1/4}$ as $b\to\infty$ 
[since $\alpha_1(1,+\infty)=2/\sqrt{3}$ and $\beta_1(1,+\infty)=0$]. Figure 1 contains two additional curves (see the inlaid box), 
illustrating the accuracy of two expressions approximating the ratio $T_b/T_{eff}$: one is Eddington's formula $T_b/T_{eff}=
(1+3b/4)^{1/4}$, which is derived e.g. in \cite{goody-yung1989}, p. 392, the other has been developed by one of us in Ref. \cite
{pelkowski1993} [equation preceding Eq. (39)] using a variational approach. We note that the accuracy of the latter approximation 
is better than $10^{-4}$ whatever $b\geq 0$.\\
\begin{figure}[htb]
\begin{center}
\resizebox{0.9\hsize}{!}{\includegraphics{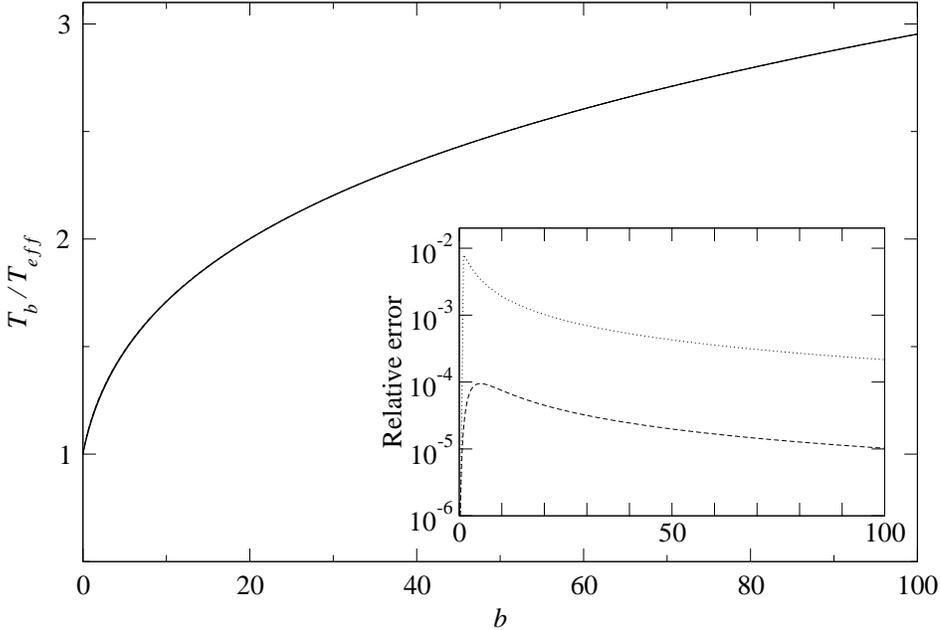}}
\end{center}
\caption{$T_b/T_{eff}$ versus optical thickness $b$. In the box: dotted and dashed curves represent the accuracy of two approximate expressions of $T_b/T_{eff}$: Eddington's and Pelkowski's, respectively.}
\label{fig1}
\end{figure}
When the atmosphere is in LTE, one can derive its temperature $T(b,\tau)$ by remarking that
\begin{equation}
\label{eq83}
S(b,\tau)=B[T(b,\tau)]=(\sigma/\pi)T^4(b,\tau),
\end{equation}
so that, from Eq. (\ref{eq75})
\begin{equation}
\label{eq84}
\frac{T(b,\tau)}{T_{b}}=[\xi_0(1,b,b-\tau)]^{1/4}=[1-\xi_0(1,b,\tau)]^{1/4}.
\end{equation}
Consequently,
\begin{equation}
\label{eq85}
T^4(b,\tau)+T^4(b,b-\tau)=T_b^4.
\end{equation}
From Eqs. (\ref{eq84}) and (\ref{eq35}), the surface values of the temperature are
\begin{equation}
\label{eq86}
\frac{T(b,0)}{T_{b}}=[\frac{1}{2}\beta_0(1,b)]^{1/4}\qquad,\qquad \frac{T(b,b)}{T_{b}}=[\frac{1}{2}\alpha_0(1,b)]^{1/4}.
\end{equation}  
The last equation shows that the temperature presents a discontinuity jump on the boundary surface $\tau=b$. This feature is well known 
to thermal engineers working on the radiative heat transfer through a gas enclosed between heated walls: see \cite{heaslet-warming1965} 
and references cited therein. It is also known in the context of thermal radiation in planetary atmospheres, see for example 
\cite{goody-yung1989} and \cite{pelkowski1993}. We note that this temperature slip at $\tau=b$ is only due to the finite 
character of the slab and has nothing to do with the hypotheses added in this section: grey atmosphere, radiative equilibrium, LTE. This can be seen by calculating $S(b)$ from Eq. (\ref{eq4}).
Supposing $B^*=0$, we find $S(b)/B(T_b)=(1/2)\alpha_0(a,b)$, which is somewhat more general than the second Eq. (\ref{eq86}). 
Our interpretation of this feature in a finite slab with a free surface is as follows: because of the escape of photons from the top boundary plane of the slab, 
the radiation field cannot be strictly isotropic within the slab, in particular very close to the boundary plane $\tau=b$ 
(if $b$ is finite). The radiation propagating in the direction of decreasing $\tau$ is thus anisotropic as long as it is 
calculated at level $\tau<b$, and it is ``abruptly'' isotropic at $\tau=b$ owing to the choice of the boundary conditions. 
This originates a discontinuity in the intensity integrated over incoming directions, hence in the mean intensity, the 
source function and the temperature.\\
The above results are useful for the modeling of planetary and stellar atmospheres, because models are always constructed 
in an iterative way, and a first temperature distribution is necessary to start the iterations. Many modern codes use the 
Hopf distribution
\begin{equation}
\label{eq87}
\frac{T(\tau)}{T_{eff}}=\{\frac{3}{4}[\tau+q(\tau)]\}^{1/4},
\end{equation}
where $q$ is Hopf's function \cite{ivanov1973}
\begin{equation}
\label{eq88}
q(\tau)=\frac{1}{\sqrt{3}} \left\{ 1+\frac{\tau}{2}\int_0^1 \frac{g(1,u)}{H(1,u)} \, \gamma^*(1,\tau/u) \frac{\d u}{u} \right\}.
\end{equation}
In an atmosphere code, this function is currently approximated by 2/3 whatever $\tau$, and Hopf's relation becomes Eddington's law.
 The exact temperature distribution (\ref{eq87}) was derived by Hopf \cite{hopf1934} for a semi-infinite, grey atmosphere 
in both radiative and local thermodynamical equilibrium, with no internal sources except at infinity. Equation (\ref{eq84})
leads to its generalization to a finite atmosphere illuminated by a black body of temperature $T_b$. The latter is connected with the 
effective temperature of the atmosphere via Eq. (\ref{eq82}). Eliminating $T_b$ from Eqs. (\ref{eq82}) and (\ref{eq84}),
one obtains
\begin{equation}
\label{eq89}
\frac{T(b,\tau)}{T_{eff}}=\left\{\frac{1-\xi_0(1,b,\tau)}{\beta_0(1,b)(\alpha_1+\beta_1)(1,b)}\right\}^{\frac{1}{4}}, 
\end{equation}
or, from Eq. (\ref{eq61}),
\begin{equation}
\label{eq90}
\frac{T(b,\tau)}{T_{eff}}=\left\{\frac{1}{2} \frac{\psi_-(1,b,\tau)}{(\alpha_1+\beta_1)(1,b)}\right\}^{\frac{1}{4}}.
\end{equation}
As $b \to +\infty$, $\psi_-(1,b,\tau) \to \psi(1,+\infty,\tau) = \sqrt{3}[\tau+q(\tau)]$ from Eqs. (\ref{eq65}), (\ref{eq55}) 
and (\ref{eq88}), $\alpha_1(1,b) \to \alpha_1(1)=2/\sqrt{3}$, $\beta_1(1,b) \to 0$ and Eq. (\ref{eq87}) is retrieved. 
We thus have $T(+\infty,\tau)=T(\tau)$ and Eq. (\ref{eq90}), with $\psi_{-}(1,b,\tau)$ given by (\ref{eq59}), generalizes
 Hopf's solution (\ref{eq87}) to a finite atmosphere.\\
We have plotted in Fig. 2 the function $\tau \to T(b,\tau) / T_{eff}$ for $b=0.01,$ 0.1, 1 and $+\infty$, and observe
that $T(b,\tau)$ is very close to $T(\tau)$ whenever $b \geq 1$. 

\begin{figure}[htb]
\begin{center}
\resizebox{0.9\hsize}{!}{\includegraphics{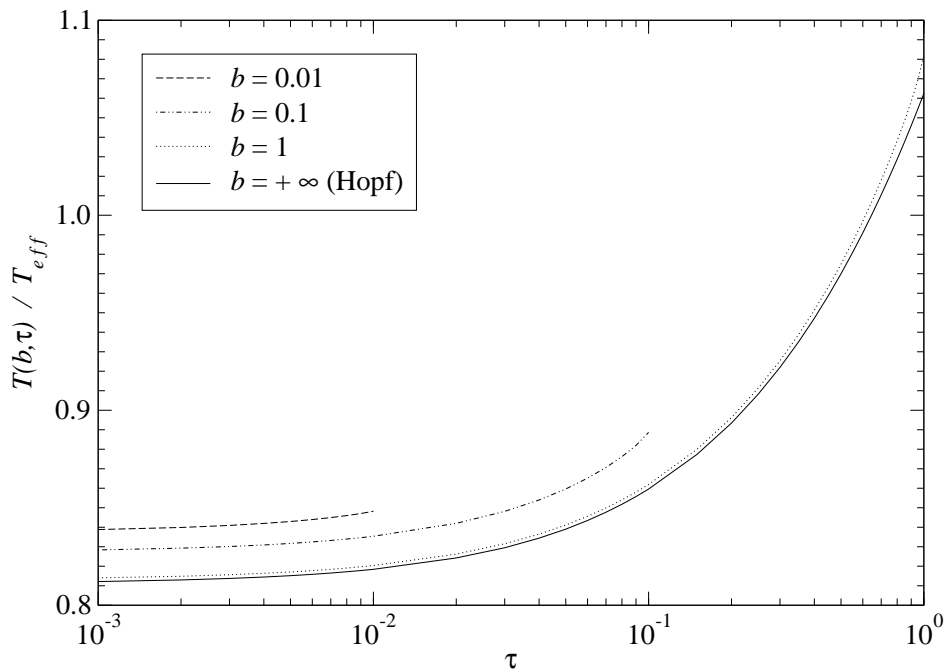}}
\end{center}
\caption{$T(b,\tau) / T_{eff}$ versus $\tau$ for $b = 0.01, 0.1, 1, +\infty$.}
\label{fig2}
\end{figure}

\section{Conclusion}\label{sec8}
The problem of a slab irradiated by an isotropic radiation field is by now a classical one. In the present article, its 
solution is expressed for unspecified values of $a$ and $b$, without thereby excluding values of astrophysical interest ($a\sim 1$, $b\gg 1$).
We intend to generalize this solution to boundary conditions of the form $I^+(\mu) = \mu^n\quad (n\geq 0)$, since in the theory of stellar 
atmospheres the boundary conditions on the surface $\tau=b$ are weakly anisotropic (diffusion approximation).\\
The conservative case is particularly interesting for the modeling of planetary and stellar atmospheres, because it is equivalent to finding 
the integrated source function of a grey atmosphere in radiative equilibrium. It thus provides a good test for models which do not presuppose 
LTE. It is also useful when the LTE condition holds, since it provides the temperature distribution (\ref{eq90}) that 
generalizes, for a finite slab, Hopf's temperature distribution (\ref{eq87}). 
%

\appendix
\renewcommand{\theequation}{\Alph{section}\arabic{equation}}
\setcounter{equation}{0}
\setcounter{section}{1}

\section*{Appendix. Proof of the relation (\ref{eq80})}

We first prove the following lemma: for any $u\neq0$
\begin{equation}
\label{eqA1}
\frac{a}{2}\int_{0}^{b}E_2(\tau)B(a,b,\tau,u)\d \tau=u[1-\tilde{\alpha}_0(a,b)X(a,b,u)-\tilde{\beta}_0(a,b)Y(a,b,u)].
\end{equation}
Since $E_2(\tau)=\int_{0}^{1}\exp(-\tau/v)\d v$, the left-hand side L of the above equation is
\begin{align}
L=&\frac{a}{2}\int_{0}^{b}\int_{0}^{1}\exp(-\tau/v)\d v B(a,b,\tau,u)\d\tau,\nonumber \\=&\frac{a}{2}\int_{0}^{1}\int_{0}^{b}B(a,b,\tau,u)\exp(-\tau/v)\d \tau \d v,\nonumber \\
=&\frac{a}{2}\int_{0}^{1}\overline{B}(a,b,1/v,u)\d v,\nonumber
\end{align}
where $\overline{B}(a,b,1/v,u)$ is the finite Laplace transform, at $1/v$, of the function $\tau \rightarrow B(a,b,\tau,u)$.
It is known (see the theorem 37.1 of \cite{busbridge1960}) that 
\begin{equation}
\label{eqA2}
\overline{B}(a,b,1/v,u)= \frac{X(a,b,u)X(a,b,v)-Y(a,b,u)Y(a,b,v)}{u+v}u v.
\end{equation}
The left-hand side of Eq. (\ref{eqA1}) is then
\begin{align}
L&=X(a,b,u)\frac{a}{2}u\int_{0}^{1}X(a,b,v)\frac{v \d v}{u+v}-Y(a,b,u)\frac{a}{2}u\int_{0}^{1}Y(a,b,v)\frac{v \d v}{u+v},\nonumber\\
&=\frac{a}{2} u X(a,b,u)[\alpha_0(a,b)-\frac{2}{a}\xi_X(a,b,u)]\nonumber\\
&\hspace{3em}-\frac{a}{2}u Y(a,b,u)[\beta_0(a,b)-\frac{2}{a}\xi_Y(a,b,u)]\nonumber
\end{align}
if we remember the definitions (\ref{eq13})-(\ref{eq15}) of $\xi_X$, $\xi_Y$, $\alpha_0$ and $\beta_0$. Equation (\ref{eqA1})
is finally derived thanks to the classical $X$-equation \cite{busbridge1960}
\begin{equation}
\label{eqA3}
X(a,b,u)\xi_X(a,b,u)-Y(a,b,u)\xi_Y(a,b,u)=X(a,b,u)-1.
\end{equation}
Integrating both members of Eq. (\ref{eqA1}) over $u$ from 0 to 1 and taking into account Eqs. (\ref{eq15}) and (\ref{eq32}),
one readily derives Eq. (\ref{eq80}).

\section*{Acknowledgements}
The authors wish to acknowledge Prof. C.E. Siewert for having pointed out to them his article with J.T. Kriese \cite{kriese-siewert1970}.

\end{document}